\pdfoutput=1
\documentclass[aps,pra,reprint,groupedaddress]{revtex4-2}
\usepackage{amsmath,amsthm,amssymb,bm,bbm}
\usepackage{graphicx}
\usepackage{mathtools}
\usepackage{enumerate}
\usepackage{dcolumn}
\usepackage{siunitx}
\usepackage{physics}
\usepackage[caption=false]{subfig}
\usepackage{placeins}
\usepackage[colorlinks,linkcolor=red,anchorcolor=blue,citecolor=green, pdfencoding=auto, psdextra]{hyperref}
\usepackage{cleveref}

\crefname{equation}{Eq.}{Eqs.}
\Crefname{figure}{Fig.}{Figs.}

\makeatletter
\renewcommand{\fnum@figure}[1]{\figurename~\thefigure.}
\makeatother

\begin{document}

\title{Preparing highly entangled states of nanodiamond rotation and NV center spin}



\author{Wen-Liang Li}

\affiliation{Institute of Physics, Beijing National Laboratory for
  Condensed Matter Physics,\\Chinese Academy of Sciences, Beijing
  100190, China}

\affiliation{School of Physical Sciences, University of Chinese
  Academy of Sciences, Beijing 100049, China}

\author{D. L. Zhou} \email[]{zhoudl72@iphy.ac.cn}

\affiliation{Institute of Physics, Beijing National Laboratory for
  Condensed Matter Physics,\\Chinese Academy of Sciences, Beijing
  100190, China}

\affiliation{School of Physical Sciences, University of Chinese
  Academy of Sciences, Beijing 100049, China}

\date{\today}

\begin{abstract}
A nanodiamond with an embedded nitrogen-vacancy (NV) center is one of the experimental systems that can be coherently manipulated within current technologies. Entanglement between NV center electron spin and mechanical rotation of the nanodiamond plays a fundamental role in building a quantum network connecting these microscopic and mesoscopic degrees of motions. Here we present a protocol to asymptotically prepare a highly entangled state of the total quantum angular momentum and electron spin by adiabatically boosting the external magnetic field.
\end{abstract}


\maketitle

\section{Introduction \label{sec:intro}}

Experimental accomplishments of cooling and controlling of microscale particles make it possible to exploit macroscopic quantum systems. The nitrogen-vacancy (NV) centers in diamond have shown impressive applications in quantum sensing, and in quantum information processing and communications~\cite{doherty2013a,chu2017c,awschalom2018}. Nanodiamonds with NV centers trapped in vacuum can be cooled into their centre-of-mass ground state~\cite{gieseler2012,delic2020,tebbenjohanns2021} and be used to generate spatial quantum superpositions~\cite{yin2013,yin2015,wan2016,pedernales2020}. While in recent years the rotation control of nanoparticle with ultra-high precision~\cite{arita2013,hoang2016,kuhn2017,kuhn2017a,rashid2018} opens the path to observing and testing rotational superpositions~\cite{delord2017a, delord2020, stickler2018, stickler2021,perdriat2022, rusconi2022}. In a view of quantum information, the coupling of NV center spin and the nanodiamond rotation contains entanglement resource~\cite{chitambar2019}. Study of the entanglement property of the spin-rotation coupled system may have potential use in quantum sensing and in quantum network. 

In this paper, we simplify the system to an ideal model by considering the nanodiamond only in an external magnetic field. The nanodiamond is treated as a rigid body and its rotation can be described by angular momentum theory in quantum physics~\cite{biedenharn1981,landau2013,yamanouchi2012}.  We show that by boosting the external magnetic field strength a highly entangled state of NV center spin and total angular momentum can be realized asymptotically.

\section{Our Model and problem \label{sec:model}}

\begin{figure}[!htp]
  \centering
  \includegraphics[width=0.36\textwidth]{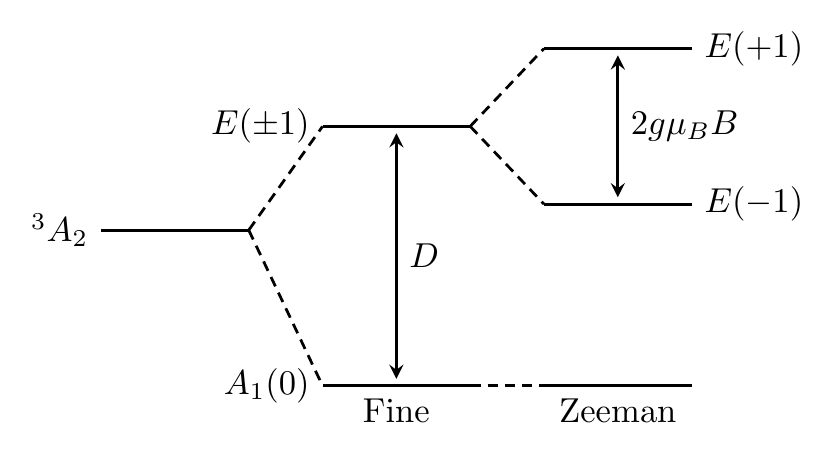}
  \caption{The fine and Zeeman structures of the NV$^{-}$ center ground state. The structure levels are denoted by their spin-orbit symmetry and spin projections \(m \in \{0,\pm 1\}\). \label{fig:e_level}}
  \vspace{-1em}
\end{figure}

\begin{figure}[!htp]
  \centering
  \includegraphics[width=0.32\textwidth]{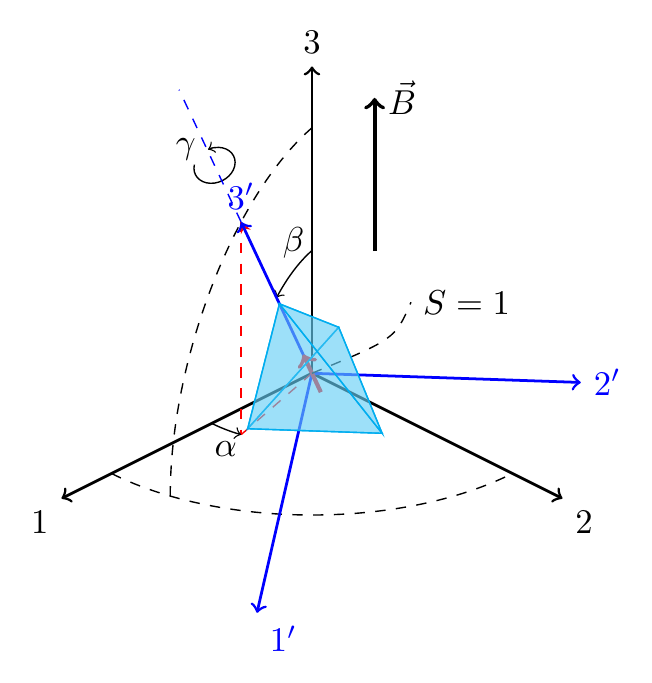}
  \vspace{-1em}
  \caption{Sketch of the nanodiamond with an embedded NV$^{-}$ center in external magnetic field \(B\). \(\{\alpha, \beta, \gamma\}\) are Euler angles between the space-fixed frame and the body-fixed frame. \label{fig:model}}  
\end{figure}

As shown in \cref{fig:model}, we consider a nanodiamond, modeled as a symmetric top whose shape is a tetrahedron. The nanodiamond hosts a single negatively charged nitrogen-vacancy center (NV\(^{-}\)) with the spin angular momentum \(\hat{\vb*{S}}\) whose quantization axis aligned with the nanodiamond symmetric axis. The ground state structure of the spin-1 NV$^{-}$ center is shown in \cref{fig:e_level}. We suppose the nanodiamond's mechanic rotation is free. In a magnetic field $\vec{B}=B \vec{e}_{3}$ along the $\vec{e}_3$-direction of the space-fixed frame with axes $\{\vec{e}_1, \vec{e}_2, \vec{e}_3\}$, the spin of the NV$^{-}$ center and  the rotation of the diamond are coupled, and the system is described by the Hamiltonian
\begin{equation}
    \label{eq:full_H}
    H = H_0 + V
\end{equation}
with
\begin{align}
    \label{eq:1}
    H_0 & = \frac{D}{\hbar} \hat{S}^{\prime 2}_3  + \frac{1}{2} \frac{\hat{\vb*{L}}^2}{I_1} + \frac{1}{2}\biggl( \frac{1}{I_3} - \frac{1}{I_1} \biggr) {\hat{L}_3^{\prime 2}} , \\
    V & = \frac{g \mu_B}{\hbar} B \sum_{i = 1}^{3} \hat{S}_i^{\prime} \hat{\vec{e}}_{i}^{\prime} \vdot \vec{e}_{3}  \label{eq:V}
\end{align}
where \(\{\hat{\vec{e}}_1^\prime,\hat{\vec{e}}_2^\prime,\hat{\vec{e}}_3^\prime\} \) are the principle axes of the nanodiamond and form the body-fixed frame, which are dynamic operators related with rotations. \(\hat{\vb*{L}}\) is the rotation angular momentum operator. \(\{I_1,I_2,I_3\}\) are the principle inertia moments with \(I_1 = I_2\) for a symmetric top. \(\hat{S}'_i\) (\(\hat{L}'_i\)) is the projection component operator of \(\hat{\vb*{S}}\) (\(\hat{\vb*{L}}\)) along the body-fixed \(\vec{e}_i^{\prime}\)-axis. \(\mu_B\) is the Bohr magneton.  The zero-field-splitting of the spin triplet ground state  \(D \simeq (2\pi)\SI{2.87}{\giga\hertz}\), and the isotropic NV$^{-}$ center electron g-factor \(g \simeq 2.0028\)~\cite{loubser1978}.

When we write the Hamiltonian in body frame, we should be careful about the commutation relations of the operators~\cite{rusconi2016}. Notice that the spin compenents \(\hat{S}'_i\) are not commuting with the angular momentum compenents \(\hat{L}'_i\) since \(\hat{S}'_i \equiv \hat{\vec{e}}'_i \vdot \hat{\vb*{S}}\), we introduce the total angular momentum operator \(\hat{\vb*{J}} = \hat{\vb*{L}} + \hat{\vb*{S}}\) whose projection compenents on body frame axes are commuting with the spin compenents, i.e. \(\comm*{\hat{J}'_i}{\hat{S}'_j} = 0\). 

The problem we aim to solve is to investigate how the entanglement between nanodiamond's total angular momentum and its spin in the thermal equilibrium state varies as a function of magnetic field and temperature, where the thermal state is
\begin{equation}
    \label{eq:2}
    \rho(B,T) =\frac{e^{-\beta H}}{\mathrm{Tr} e^{-\beta H}},
\end{equation}
where $\beta=\frac{1}{k_{B}T}$ with $k_B$ being the Boltzmann constant. As the temperature limits to zero, the thermal equilibrium state becomes the ground state:
\begin{equation}
    \label{eq:3}
    \lim_{T\rightarrow 0} \rho(B, T) = |G(B)\rangle \langle G(B)|.
\end{equation}


\section{Eigen problem of the Hamiltonian \label{sec:eigen-problem}}

Before exploring the entanglement between the total angular momentum and the spin in the thermal equilibrium state $\rho(B,T)$ or in the ground state $|G(B)\rangle$, it is necessary for us to solve the eigen problem of the Hamiltonian~\eqref{eq:full_H}. Rewrite the Hamiltonian by inserting \(\hat{\vb*{L}} = \hat{\vb*{J}} - \hat{\vb*{S}} = \hat{\vb*{J}} + \hat{\vb*{K}}\), 
\begin{align}
  H_0 &= \frac{D}{\hbar} \hat{K}^{\prime 2}_3  + \frac{1}{2 I_1} (\hat{\vb{J}}^2 + 2 \hat{J}'_3 \hat{K}'_3 + \hat{J}'_+ \hat{K}'_- + \hat{J}'_- \hat{K}'_+) \notag \\
  &+ \pqty{\frac{1}{I_3} - \frac{1}{I_1}} \frac{ (\hat{J}'_3 + \hat{K}'_3)^{2}}{2} , \notag \\
  V &= - \frac{g \mu_B}{\hbar} \hat{\vb{K}} \vdot  \vb{B} , 
\end{align}
where we have let \(\hat{\vb*{K}} \equiv - \hat{\vb*{S}}\) with \(K \equiv S \equiv 1\) in our problem and dropped \(\hat{\vb{S}}^2=S(S+1)\). And the ladder operators are 
\begin{align}
  \hat{J}'_{\pm} & = \hat{J}'_1 \mp \mathrm{i} \hat{J}'_2 , \\
    \hat{K}'_{\pm} & = \hat{K}'_1 \mp \mathrm{i} \hat{K}'_2 .
\end{align}

\subsection{Basis states based on \(H_0\)}
\label{sec:basis-states-based}

First we study the degree of the nanodiamond rotation. Because \(\hat{\boldsymbol{L}}\) is an angular momentum operator, it obeys the following commutation relations
\begin{align}
  \label{eq:4}
  \comm{\hat{L}_{i}}{\hat{L}_{j}} & = i\hbar \epsilon_{ijk}\hat{L}_{k}, \\
  \comm{\hat{L}_{i}}{\hat{\vec{e}}_{l}^{\prime} \vdot \vec{e}_{j}} & = i\hbar \epsilon_{ijk}  \hat{\vec{e}}_{l}^{\prime} \vdot \vec{e}_{k},
\end{align}
where $i,j,k\in\{1,2,3\}$ and $l\in\{1,2,3\}$, and $\epsilon_{ijk}$ is antisymmetric tensor with $\epsilon_{123}=1$. Note that $\hat{L}'_i\equiv \hat{\vec{e}}'_i\vdot \hat{\vb*{L}}$. Then a direct calculation gives
\begin{equation}
  \label{eq:5}
  \comm{\hat{L}'_i}{\hat{L}'_j}=-i\hbar \epsilon_{ijk}\hat{L}'_{k},
\end{equation}
where $i,j,k\in\{1,2,3\}$. Compared \cref{eq:5} with \cref{eq:4}, it is interesting to observe that a negative sign appears in the commutation relations of the angular momentum in the body-fixed frame. With the relations \(\hat{S}'_i \equiv \hat{\vec{e}}'_i \vdot \hat{\vb*{S}} = \sum_{j=1}^3 \hat{\vec{e}}'_i \vdot \vec{e}_j \hat{S}_j\), one can give the following commutation relations: 
\begin{align}
  \comm{\hat{J}'_i}{\hat{J}'_j} &= - \mathrm{i} \epsilon_{ijk} \hat{J}'_k , \\
  \comm{\hat{K}'_i}{\hat{K}'_j} &= - \mathrm{i} \epsilon_{ijk} \hat{K}'_k , \\
  \comm{\hat{J}'_3}{\hat{K}'_3} &= 0 ,\\
  \comm{\hat{J}_3}{\hat{K}'_3} &= 0 .
\end{align}
The total angular momentum
\begin{equation}
  \label{eq:6}
  \hat{\vb*{J}}^2 = \sum_{i} \hat{J}_{i}^2 = \sum_i \hat{J}_i^{' 2},
\end{equation}
which builds a relation between the angular momentum operators in these two frames. Then we find that
\begin{align}
  \label{eq:7}
  \comm{\hat{\vb*{J}}^2}{\hat{J}_3} & = 0, \\
  \comm{\hat{\vb*{J}}^2}{\hat{J}'_3} & = 0, \\
  \comm{\hat{J}_3}{\hat{J}'_3} & = 0,
\end{align}
which implies that $\{ \hat{\vb*{J}}^2, \hat{J}_3, \hat{J}'_3,  \hat{\vb*{K}}^2, \hat{K}'_3\}$ forms a complete set of commuting observables for the degree of freedom of the system. They have the common eigenstates
\begin{equation} \label{eq:phi}
  \ket{\varphi} = \ket{ J m_J k_J , K k_K} .
\end{equation}
such that
\begin{align}
  \hat{\vb*{J}}^2 \ket{\varphi} &= J(J+1) \ket{\varphi},\\
  \hat{J}_3 \ket{\varphi} &= m_J \ket{\varphi},\\
  \hat{J}'_3 \ket{\varphi} &= k_J \ket{\varphi}, \\
  \hat{\vb*{K}}^2 \ket{\varphi} &= K(K+1) \ket{\varphi}, \\
  \hat{K}'_3 \ket{\varphi} &= k_K \ket{\varphi},
\end{align}
where $J\in\{0,1,\cdots\}$, $m_J,k_J\in\{-J,-J+1,\cdots,J-1,J\}$, \(k_K \in \{-1,0,1\}\), and these common eigenstates form a set of base vectors of the full Hilbert space. Then the matrix elements of the ladder operators \(\hat{J}'_{\pm}\) and \(\hat{K}'_{\pm}\) are given by the following relations, 
\begin{align}
  &\hat{J}'_{\pm} \ket{J m_J k_J,K k_K} \notag\\
  &= \sqrt{(J \mp k_J)(J \pm k_J +1)} \ket{J m_J k_J \pm 1, K k_K} , \\
  &\hat{K}'_{\pm} \ket{J m_J k_J,K k_K} \notag \\
  &= \sqrt{(K \mp k_K)(K \pm k_K +1)}  \ket{J m_J k_J, K k_K \pm 1} .
\end{align}
For \(K = 1\), it's convenient to write out the \(\hat{K}'_i\) matrices
\begin{align}
  K'_1 &= \frac{1}{\sqrt{2}}
  \begin{pmatrix}
      0 & 1 & 0 \\
      1 & 0 & 1 \\
      0 & 1 & 0
  \end{pmatrix}, \\
  K'_2 & = \frac{-i}{\sqrt{2}}
   \begin{pmatrix}
      0 & -1 & 0 \\
      1 & 0 & -1 \\
      0 & 1 & 0
  \end{pmatrix},  \\
  K'_3 &= \begin{pmatrix}
      1 & 0 & 0 \\
      0 & 0 & 0 \\
      0 & 0 & -1
  \end{pmatrix}.
\end{align}

\subsection{Analytical matrix elements in \texorpdfstring{$V$}{}}
\label{sec:analyt-matr-elem}

The next task is to calculate the matrix element \(\bra*{J' m_J' k_J' , K k_K'} V \ket*{J m_J k_J , K k_K}\). In other words, we need to calculate the matrix elements \(\bra*{J' m_J' k_J'} \hat{\vec{e}}'_i \vdot \vec{e}_j \ket*{J m_J k_J} \). 

Because all the operators $\{ \hat{\vec{e}}'_l \vdot \vec{e}_{i}\}$ are commutative to each other, we can introduce their common eigenstates
\begin{equation}
  \label{eq:12}
  \hat{\vec{e}}'_l \vdot \vec{e}_{i} |\alpha \beta \gamma\rangle = \vec{e}'_l \vdot \vec{e}_{i} |\alpha \beta \gamma\rangle,
\end{equation}
where \(\{\alpha,\beta,\gamma\}\) are three Euler angles (illustrated in \cref{fig:model}). In particular
\begin{align}
  \label{eq:13}
  \hat{\vec{e}}'_1 \vdot \vec{e}_3 |\alpha \beta \gamma\rangle & = -\cos{\gamma} \sin{\beta} |\alpha \beta \gamma\rangle,\\
  \hat{\vec{e}}'_2 \vdot \vec{e}_3 |\alpha \beta \gamma\rangle & = \sin{\gamma} \sin{\beta} |\alpha \beta \gamma\rangle,\\
  \hat{\vec{e}}'_3 \vdot \vec{e}_3 |\alpha \beta \gamma\rangle & = \cos{\beta} |\alpha \beta \gamma\rangle.
\end{align}
Then
\begin{align}
  \label{eq:14}
  &\bra{J' m_J' k_J'} \hat{\vec{e}}'_i \vdot \vec{e}_3 \ket{J m_J k_J} \notag \\
  &= \int \dd \omega \vec{e}'_i \vdot \vec{e}_3 \braket{J' m_J' k_J'}{\alpha \beta \gamma} \braket{\alpha \beta \gamma}{J m_J k_J},
\end{align}
where $\mathrm{d} \omega = \sin\beta \mathrm{d} \alpha \mathrm{d} \beta \mathrm{d} \gamma$ with $\alpha,\gamma\in[0,2\pi]$ and $\beta\in[0,\pi]$. The eigenfunction \(\braket{\alpha \beta \gamma}{J m_J k_J}\)  can be written as
\begin{align}
    \label{eq:15}
    \braket{\alpha \beta \gamma}{J m_J k_J} = \sqrt{\frac{2J+1}{8 \pi^2}} D_{k_J m_J}^{(J)} (\alpha, \beta, \gamma),
\end{align}
where \(D_{k_J m_J}^{(J)} (\alpha, \beta, \gamma)\) is the matrix representation of Euler rotation operator \(\hat{R}(\alpha,\beta,\gamma)\) in Hilbert space (more details see~\cref{app:a}). 

A set of spherical basis vectors in space-fixed frame can be defined as 
\begin{align}
    \label{eq:16_1}
    \vec{\xi }_{+1} &\equiv -(\vec{e}_1 + i \vec{e}_2)/\sqrt{2}, \\
    \vec{\xi }_0 &\equiv \vec{e}_3, \label{eq:16_2}\\
    \vec{\xi }_{-1} &\equiv (\vec{e}_1 - i \vec{e}_2)/\sqrt{2},\label{eq:16_3}
\end{align}
with the orthogonality relations
\begin{equation}
    \label{eq:17}
    \vec{\xi}^*_\mu \vdot \vec{\xi}_\nu = \delta_{\mu \nu},
\end{equation}
where \(\nu, \mu \in \{+1,0,-1\}\), and $\vec{\xi}^{\ast}_{\mu}$ is the complex conjugate of $\vec{\xi}_{\mu}$.
Then we can show that $\{\vec{\xi}_{\mu}\}$  and \(\{\vec{\xi}'_\mu\}\) obey the transformation
\begin{equation}
    \label{eq:18}
    \vec{\xi} _\mu = \sum_\nu D^{(1)}_{\nu \mu} (\alpha, \beta, \gamma) \vec{\xi}'_\nu .
\end{equation}
Combine \cref{eq:16_1,eq:16_2,eq:16_3,eq:17,eq:18}, we have 
\begin{equation}
    \label{eq:19}
    \vec{\xi}^{\prime *}_{\mu} \vdot \vec{e}_3 = \vec{\xi}^{\prime *}_{\mu } \vdot \vec{\xi}_0 = D^{(1)} _{\mu 0} (\alpha, \beta, \gamma) .
\end{equation}
Then 
\begin{align}
    \label{eq:20}
    \vec{e}^{\prime}_1 \vdot \vec{e}_3 &=  \pqty{- D^{(1)}_{1 0}(\alpha,\beta,\gamma) + D^{(1)}_{-1\;0}}/ \sqrt{2} , \notag \\
    \vec{e}^{\prime}_2 \vdot \vec{e}_3 &=  - i \pqty{ D^{(1)}_{1 0}(\alpha,\beta,\gamma) + D^{(1)}_{-1\;0}}/ \sqrt{2} , \notag \\
    \vec{e}^{\prime}_3 \vdot \vec{e}_3 &= D^{(1)}_{0 0}(\alpha,\beta,\gamma).
\end{align}
Employing the identity of the integral of three \(D_{\nu \mu }^{(J)} (\alpha, \beta, \gamma)\) functions:
\begin{align} 
    \label{eq:21}
    & \int D_{m_1' m_1 }^{(J_1)} (\alpha, \beta, \gamma) D_{m_2' m_2 }^{(J_2)} (\alpha, \beta, \gamma) D_{m_3' m_3 }^{(J_3)} (\alpha, \beta, \gamma) \frac{\dd \omega}{8 \pi^2} \notag \\
    & = \begin{pmatrix}
        J_1 & J_2 & J_3 \\
        m'_1 & m'_2 & m'_3
        \end{pmatrix} 
            \begin{pmatrix}
            J_1 & J_2 & J_3 \\
            m_1 & m_2 & m_3
            \end{pmatrix} ,
\end{align} 
and their complex conjugate relations 
\begin{equation}
    \label{eq:22}
    D_{\nu \mu }^{(J) *} (\alpha, \beta, \gamma) = (-1)^{\nu - \mu} D_{-\nu, -\mu }^{(J)} (\alpha, \beta, \gamma) ,
\end{equation}
we arrive at an analytical result of the integral
\begin{align}
    \label{eq:23}
    &\bra{J' m_J' k_J'} D_{\mu 0}^{(1)} (\alpha, \beta, \gamma) \ket{J m_J k_J} \notag \\
    &= \int \frac{\dd \omega}{8 \pi ^2} \sqrt{(2J'+1)(2J+1)} \notag \\
    & \quad \times D_{k_J' m_J' }^{(J') *} (\alpha, \beta, \gamma) D_{\mu 0}^{(1)} (\alpha, \beta, \gamma) D_{k_J m_J }^{(J)} (\alpha, \beta, \gamma) \notag \\
    &= (-1)^{k_J' - m_J'} \sqrt{(2J'+1)(2J+1)} \notag \\
    & \quad \times  \begin{pmatrix} 
        J' & 1 & J \\
        -k_J' & \mu & k_J
        \end{pmatrix} 
            \begin{pmatrix}
            J' & 1 & J \\
            -m_J' & 0 & m_J
            \end{pmatrix} ,
\end{align}
where the rhs of \cref{eq:21} are two \(3J\)-symbols. Inserting \cref{eq:23} and \cref{eq:20} into \cref{eq:14}, we get the matrix element \(\bra{J' m_J' k_J'} \hat{\vec{e}}'_i \vdot \vec{e}_3 \ket{J m_J k_J}\). Combine with \(\hat{K}'_i\), we get the matrix element \(\bra*{J' m_J' k_J' , K k_K'} V \ket*{J m_J k_J , K k_K}\). 

\begin{figure*}[!htb]
  \vspace{-1em}
  \centering
  \subfloat[]{
      \centering
      \label{fig:3a}
      \includegraphics[width=.32\textwidth]{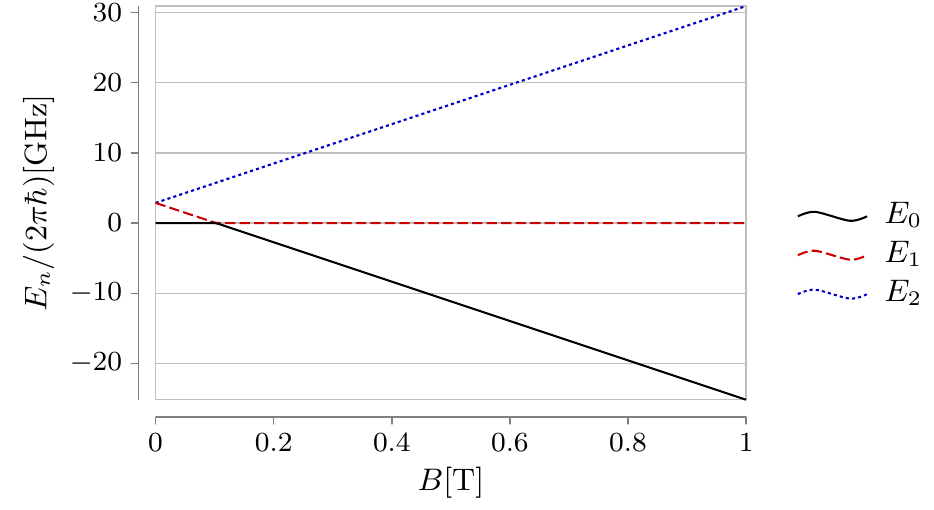}
  }
  \hspace{-1em}
  \subfloat[]{
      \centering
  \label{fig:3b}
  \includegraphics[width=.32\textwidth]{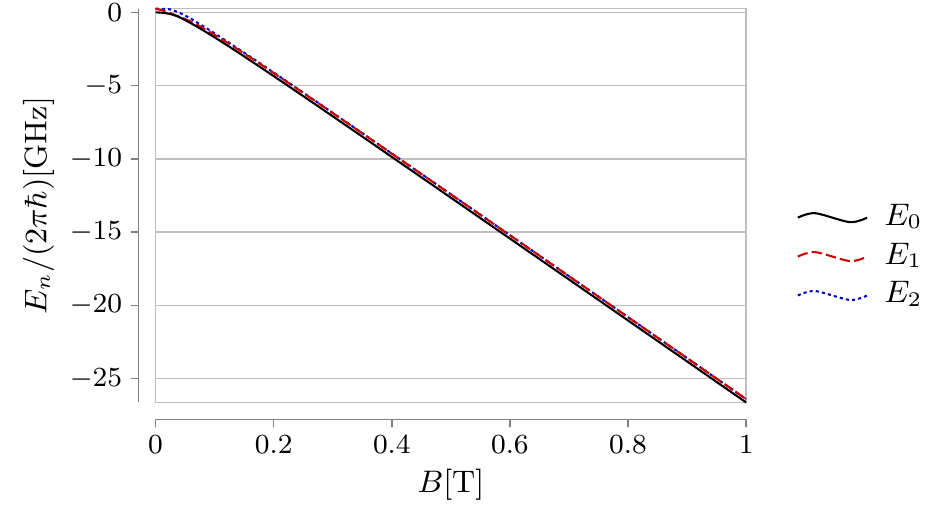}
  }
  \hspace{-1em}
  \subfloat[]{
      \centering
  \label{fig:3c}
  \includegraphics[width=.32\textwidth]{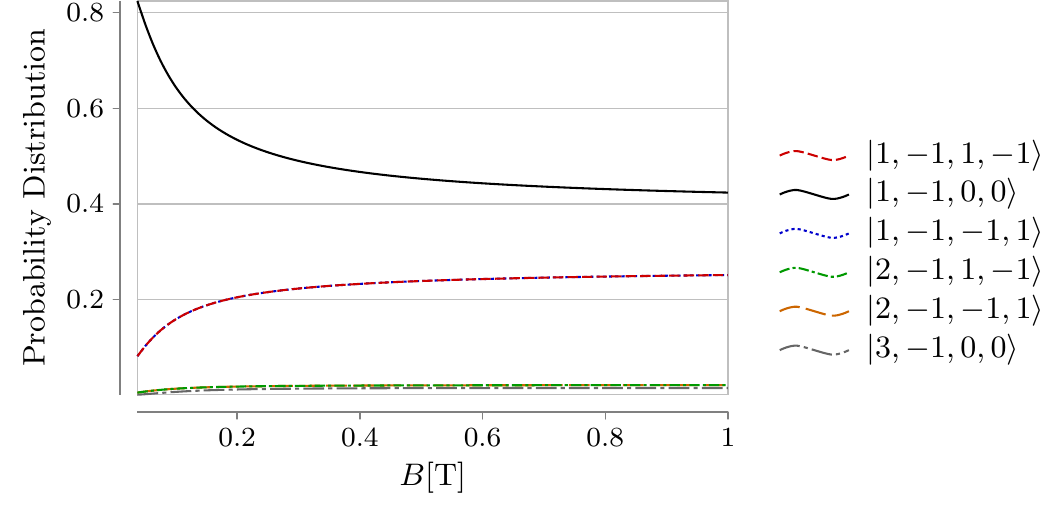}
  }
  \caption[]{\label{fig:3} \subref{fig:3a} The energy levels of the spin Hamiltonian changing with the external magnetic field \(B\). \subref{fig:3b} The energy levels of the full spin-rotation Hamiltonian in \cref{eq:full_H} with \(J_{\mathrm{max}} = 4\). \subref{fig:3c} The main probability distribution of the ground state on base kets \(\ket{J m_J k_J k_K}\) with \(J_{\mathrm{max}} = 4\). }
  \vspace{-1em}
\end{figure*}

\subsection{Numerical results on eigen energies}
\label{sec:numer-results-eigen}

Before numerical solving the eigen problem of Hamiltonian $H$, we first need to give the values of the inertia momentum $\{I_1,I_2, I_3\}$, which is determined by the nanodiamond size. In our calculations, we takes the bottom side length of the nanodiamond $a = \SI{1}{\nm}$ and height \(h = \SI{1.225}{\nm}\), which leads to $I_1=\SI{5.06e-44}{kg.m^{2}}$ and $I_3=\SI{3.11e-44}{kg.m^{2}}$.

Since we focus on low energy physics of our system, it is nature to introduce a cutoff via a maximum angular momentum $J_{\text{max}}$ in our numerical calculations. To ensure the convergence of our physical results, we set \(J_\textrm{max} = 4\) (convergence tests see~\cref{app:b}). Then we solve the eigen equation of full Hamiltonian
\begin{equation}\label{eq:eigen_eq}
    H | \Psi_i \rangle = E_i |\Psi_i \rangle ,
\end{equation}
where 
\begin{equation}
    |\Psi_i \rangle = \sum_{J m_J k_J k_K}^{J \le J_{\text{max}}} c^{(i)}_{J m_J k_J k_K} \ket{ J m_J k_J , K k_K} .
\end{equation}
The energy levels are shown in \cref{fig:3b}.  As a comparison, \cref{fig:3a} shows the energy levels of the effective spin Hamiltonian
\begin{equation}
    H_S = D \hat{S}_z^2 + g \mu_B B \hat{{S_z}}, \label{eq:Hs}
\end{equation}
which generally describes a resting NV$^{-}$ center with magnetic field \(\vec{B} = B \vec{e}_z\) in NV$^{-}$ axis.

From \cref{fig:3a} and \cref{fig:3b}, we observe that in different magnetic fields, the energies of the ground state for our system are similar as those of the Hamiltonian without considering the rotation given by \cref{eq:Hs}. However, our ground states become highly entangled states which mainly involve six components as shown in \cref{fig:3c}.

\section{Entanglement of thermal equilibrium state}
\label{sec:entangl-therm-state}

When our quantum system interacts with its thermal environment, it will finally arrives at a steady state: the thermal equilibrium state. Now we are ready to study the entanglement properties in these thermal equilibrium states, which will be useful to guide us to provide a natural protocol to prepare entanglement between nanodiamond rotation and NV$^{-}$ center spin.

\subsection{Entanglement of ground states}

First we study the entanglement properties of the ground state, i.e., the thermal equilibrium state when the temperature limits to zero. For a ground state $\ket{G(B)}_{JS}$ at magnetic field $B$, the entanglement entropy is defined as:
\begin{equation}
    S(\rho_S(B)) = -\Tr_S \rho_S(B) \log_2(\rho_S(B)),
\end{equation}
where \(\rho_S(B)\) is the reduced spin density matrix of \(\ket{G(B)}_{JS}\). Because the dimension of the Hilbert space of NV$^{-}$ center spin $d_S=3$, the entanglement entropy $S(\rho_S)\le\log_23$, where the equality is taken if and only if the ground state $\ket{G(B)}_{JS}$ is maximally entangled.

Numerical results on the entanglement entropy $S(\rho(B))$ are shown in \cref{fig:4a}. With the increasing of magnetic field $B$, the entanglement of the ground state grows from $0$ to approximately $\log_23$, which implies that the ground state limits to a highly entangled state in a large magnetic field $B$.

\subsection{Entanglement of thermal equilibrium states at low temperatures}

\begin{figure*}[!htp]
  \vspace{-1em}
  \centering
  \subfloat[]{
      \centering
      \label{fig:4a}
      \includegraphics[width=.32\textwidth]{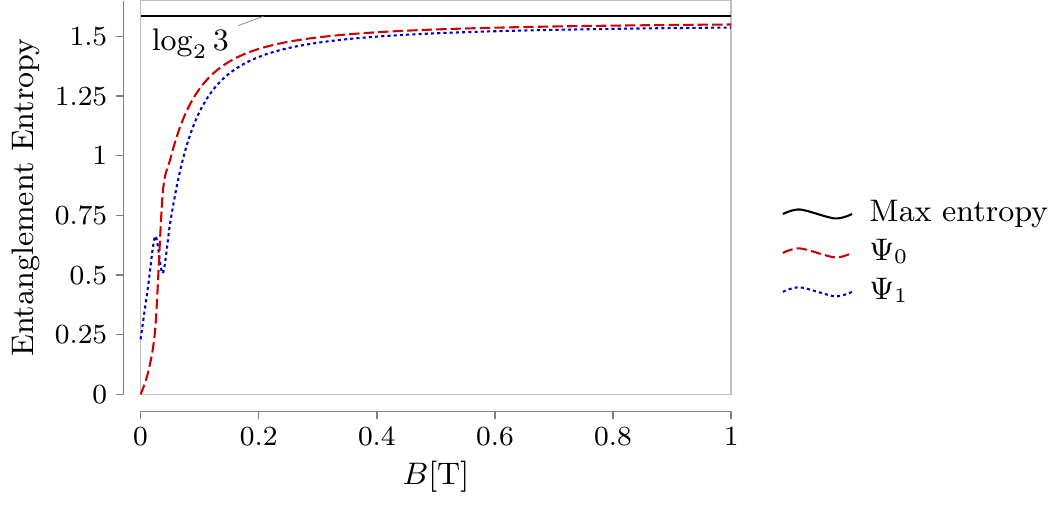}
  }\hspace{-1em}
  \subfloat[]{
      \centering
      \label{fig:4b}
      \includegraphics[width=.32\textwidth]{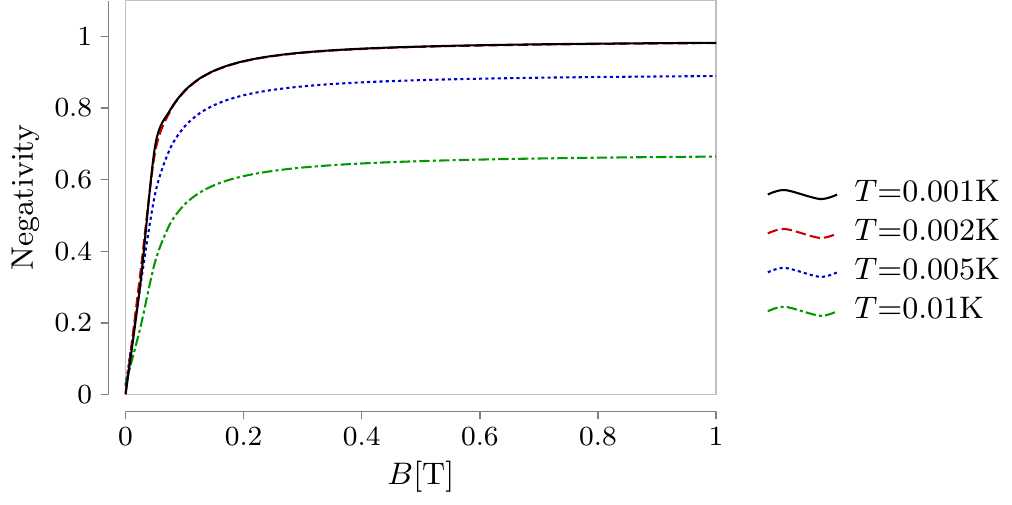}
  }\hspace{-1em}
  \subfloat[]{
          \centering
      \label{fig:4c}
      \includegraphics[width=.32\textwidth]{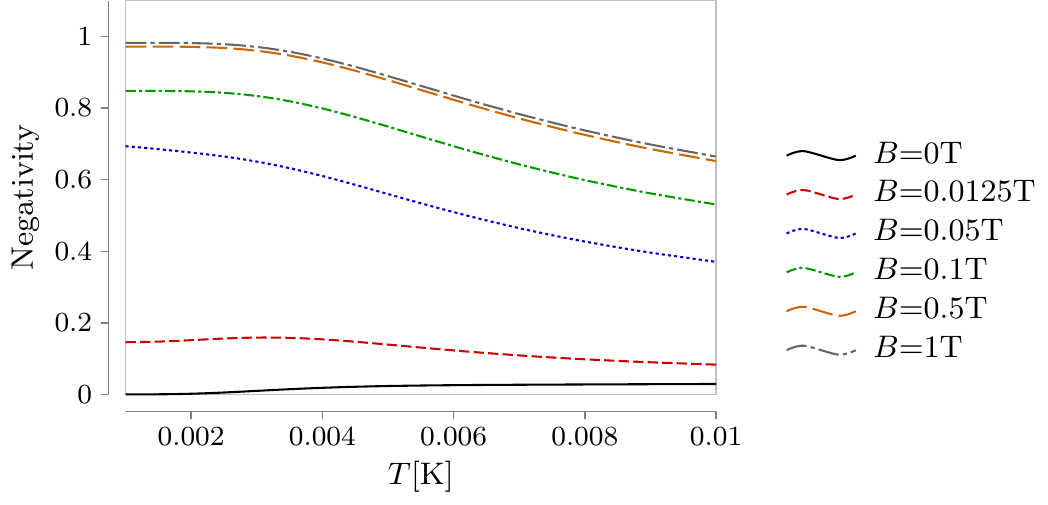}
  }
  \caption[]{\label{fig:fig_Neg} The size of the particle is about \(a = \SI[]{1}{\nm}\). \subref{fig:4a} The entanglement entropy of the ground sate and the first excited state compared with the maximum entanglement entropy for the maximum angular quantum number \(J_\textrm{max} = 4 \). The negativity of the thermal entanglement state changing with magnetic field \(B\) at some fixed temperatures \(T\) \subref{fig:4b} and with absolute temperature at some fixed magnetic fields \(B\) \subref{fig:4c}. }
  \vspace{-1em}
\end{figure*}

At temperature \(T\), the thermal equilibrium state can be represented as
\begin{equation}
    \rho_{JS} (B, T) = \frac{1}{Z} \sum_i e^{-\beta E_i} |\Psi_i\rangle \langle \Psi_i |,
\end{equation}
where the partition function
\begin{equation}
    Z = \sum_i e^{-\beta E_i} ,
\end{equation}
and \(|\Psi_i\rangle\) is the eigenvector of $H$ with eigenvalue \(E_i\), which has been obtained numerically in the previous section.

Because the thermal equilibrium state $\rho_{JS}(B, T)$ is a mixed state, its entanglement can not be characterized by the entanglement entropy $S(\rho_S)$, which is valid for characterization of entanglement for pure states. To study the entanglement property of the thermal equilibrium state, we introduce another entanglement measure, \emph{negativity}~\cite{vidal2002,horodecki2009}
\begin{equation}
    \mathcal{N} (\rho_{JS}) = \sum_{\lambda < 0} \mid \lambda \mid = \frac{\norm{\rho_{JS} ^{T_S}}_1 - 1}{2},
\end{equation}
where \(\rho_{JS} ^{T_S}\) is the partial transpose of spin index for the bipartite mixed state \(\rho_{JS}\), \(\lambda\) is the eigenvalue of \(\rho_{JS} ^{T_S}\). The negativity corresponds to the absolute value of the sum of negative eigenvalues of \(\rho_{JS} ^{T_S}\). \( \norm{\rho_{JS} ^{T_S}}_1 \) is the trace norm of \(\rho_{JS} ^{T_A}\). The negativity \(\mathcal{N}\) is a computable measure of entanglement for a mixed state and vanishes for separable states.

The numerical results of the negativity are shown in \cref{fig:fig_Neg}. It is observed in \cref{fig:4b} that for a given temperature $T$, the negativity increases asymptotically to a maximum value with increasing of the magnetic field $B$. The temperature lower, the maximal value of the negativity larger. As shown in \cref{fig:4c}, for a fixed magnetic field $B$, the negativity decreases with increasing of the absolute temperature $T$. The magnetic field larger, the negativity larger. Our numerical results show that to obtain a thermal equilibrium state highly entangled, we need to increase the magnetic field larger than $\SI{0.5}{\tesla}$ and decrease the temperature below $\SI{2}{\milli\K}$.

Based on the above numerical results, we propose a simple protocol to asymptotically prepare a highly entangled state between mechanical rotation of the nanodiamond and the electron spin of NV$^{-}$ center. First, cool down the system to below $\SI{2}{\milli\K}$ at zero or weak external magnetic field strength. Then adiabatically boost the magnetic field strength to above \(B = \SI{0.5}{\tesla}\) and keep the system still in low enough temperature. Finally in thermal equilibrium, we get the thermal equilibrium state highly entangled.

\section{Discussion and conclusion}

\begin{figure}[!h]
  \centering
  \includegraphics[width=.8\columnwidth]{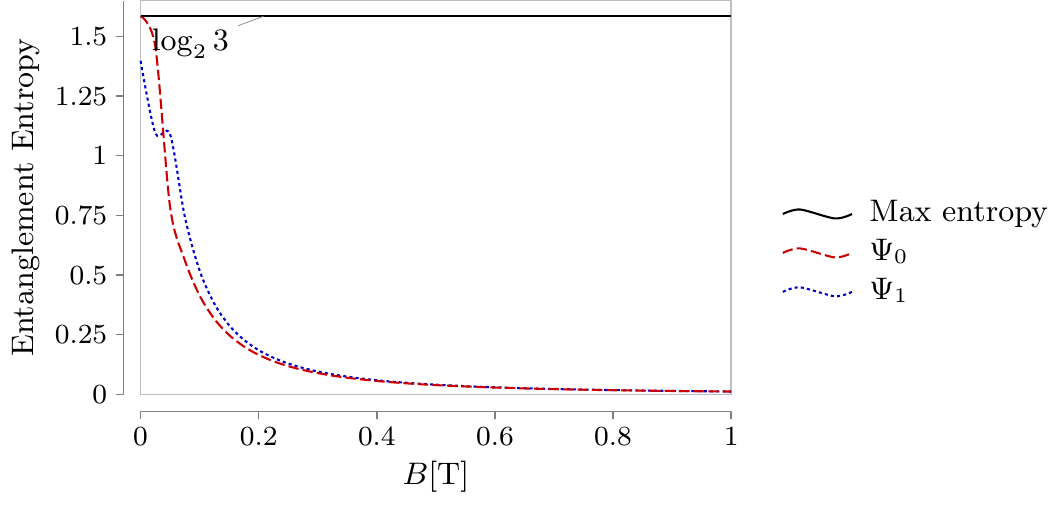}
  \caption{The entanglement entropy of the ground state and the first excited state solved in the space-fixed frame with cutoff \(L_{\mathrm{max}} = 4\). \label{fig:5} }
\end{figure}

Our model is solved in the body-fixed frame, giving different results with which solved in the space-fixed frame as shown in \cref{fig:5}. Because in space-fixed frame, the complete set of commuting observables is \(\{\hat{\vb*{L}}^2, \hat{L}_3, \hat{L}'_3,  \hat{\vb*{S}}^2, \hat{S}_3\}\) which is not all commutative with those in the body-fixed frame, i.e. \(\{ \hat{\vb*{J}}^2, \hat{J}_3, \hat{J}'_3,  \hat{\vb*{K}}^2, \hat{K}'_3\}\). This is consistent with physical interpretation that in the space-fixed frame strong enough magnetic field makes the spin occupying \(\ket{-1}\) in ground state. Then from the view point in space-fixed frame, boosting magnetic field strength just results oppsite effect--disentangelment--comparing with the view in the body-fixed frame. More details see~\cref{app:c}. 

We propose a theoretical model to describe a rotating nanodiamond with an embedded NV$^-$ center manipulated by a static external magnetic field. We neglect several factors in real experiments such as trap potential design, gravity effect, decoherence from external noises that may affect the accuracy of our model, which deserve further studies in future.

In our protocol to prepare entanglement, we propose to adiabatically boost the magnetic field strength. Theoretically, however, we do not require the boosting is adiabatic, a sudden change of  the magnetic field strength may also work after a much longer equilibrium time.

In conclusion, we explore the entanglement properties of a rotating nanodiamond with an embedded NV$^{-}$ center in an external magnetic field in a thermal equilibrium state, which includes the ground state as a special case. We find that the entanglement between nanodiamond rotation and NV\(^-\) center spin can be controlled by an external magnetic field and the temperature: larger magnetic field strength and lower temperature, more entanglement between the rotation and the spin. Our numerical results show that in our system setting when the magnetic field strength is tuned above $\SI{0.5}{\tesla}$ and the temperature is controlled below $\SI{2}{\milli\K}$, the thermal equilibrium state will be an almost maximally entangled state. Thus we propose a theoretical protocol to realize the highly entangled states of the spin-rotation coupled system asymptotically. The entanglement between the spin (a microscopic degree) and the rotation (a mesoscopic degree) is not only of interest in fundamental problems of quantum mechanics such as the border between quantum world and classical world~\cite{aspelmeyer2014}, but also may find potential use in quantum control, and in quantum sensing and in quantum network.

\begin{acknowledgments}
This work is supported by National Key Research and Development Program of China (Grant No. 2021YFA0718302 and No. 2021YFA1402104), National Natural Science Foundation of China (Grants No. 12075310), and the Strategic Priority Research Program of Chinese Academy of Sciences (Grant No. XDB28000000).
\end{acknowledgments}

\appendix
\section{\label{app:a}\(D\)-matrix and Euler rotations}
\setcounter{equation}{0}
\renewcommand\theequation{A.\arabic{equation}}

In this appendix, we give some details of the \(D-\)matrix of Euler rotations. We have chosen \(\{\vec{e}_1, \vec{e}_2, \vec{e}_3\}\) to represent the space-fixed frame and  \(\{\vec{e}'_1, \vec{e}'_2, \vec{e}'_3\}\) the body-fixed frame. In the view of passive rotations, we consider \(\vec{e}'_i\) is rotated to \(\vec{e}_i\) by rotation operator \(\hat{R}\),
\begin{equation}
  \label{eq:rotat}
  \vec{e}_i = \hat{R} \vec{e}'_i = \sum_{j=1}^3 R_{ji} \vec{e}'_j ,
\end{equation}
where 
\begin{equation}
  R_{ji} \equiv \vec{e}'_j \vdot \vec{e}_i .
\end{equation}
While in the view of active rotations, we usually define the rotation operator as \(\vec{e}'_i = \hat{Q} \vec{e}_i\) which maps a vector \(\vec{e}_i\) to a new vector \(\vec{e}'_i\) in the same frame. It is clear to see that \(\hat{R} = \hat{Q}^{-1}\) which usually gives the inverse relation of passive and active view of the same rotation transformation. In our paper, we choose the passive view on account of that we have defined two coordinate frames. And we choose Euler angles \(\{\alpha, \beta, \gamma\}\) to represent the rotation from space-fixed frame \(\{\vec{e}_1, \vec{e}_2, \vec{e}_3\}\) to body-fixed frame \(\{\vec{e}'_1, \vec{e}'_2, \vec{e}'_3\}\) which are shown in \cref{fig:model}. 

According to quantum mechanics, the generator of \(\hat{R}\) is angular momentum \(\hat{\vb*{L}}\), especially in space-fixed frame,
\begin{equation}
  \comm{\hat{L}_i}{\hat{L}_j} = i \hbar \epsilon_{ijk} \hat{L}_k ,
\end{equation}
where \(i,j,k \in \{1,2,3\}\) with \(\hat{L}_i \equiv \vec{e}_i \vdot \hat{\vb*{L}}\) and \(\epsilon_{ijk} \) is an antisymmetric tensor with \(\epsilon _{123} = 1\). Let \(\hat{D}\) be the representation of the rotation operator \(\hat{R}\) in Hilbert space, we have
\begin{equation}
  \hat{D}(\alpha,\beta,\gamma) = \exp(\frac{i \gamma \hat{L}_3}{\hbar})\exp(\frac{i \beta \hat{L}_2}{\hbar})\exp(\frac{i \alpha \hat{L}_3}{\hbar}) .
\end{equation}
And the space base kets relations are defined as 
\begin{equation}
  \hat{D}^\dagger (\alpha,\beta,\gamma) \ket{\{\vec{e}'_i\}} = \ket{\{\hat{R}\vec{e}'_i\}} = \ket{\{\vec{e}_i\}}. 
\end{equation}
In particular, for any vector operator \(\hat{\vb*{A}}\) 
\begin{equation}
  \hat{D}^\dagger \hat{\vb*{A}} \hat{D} = \hat{Q}^{-1} \hat{\vb*{A}} =\hat{R} \hat{\vb*{A}} . 
\end{equation}
Then 
\begin{align}
  \hat{D}^\dagger \hat{A}_i \hat{D}
  \notag &= \vec{e}_i \vdot \hat{D}^\dagger \hat{\vb*{A}} \hat{D} \\
  \notag &= \vec{e}_i \vdot \hat{R} \hat{\vb*{A}} \\
  \notag &= \vec{e}_i \vdot \sum_j (\hat{\vb*{A}} \vdot \vec{e}'_j) \hat{R} \vec{e}'_j \\
  \notag &= \sum_j \hat{A}'_j  \vec{e}_i \vdot \vec{e_j} \\
  &= \hat{A}'_i .
\end{align}
Since in the space-fixed frame, \(\hat{\vec{e}}'_l\) is a vector operator, 
\begin{equation}
  \comm{\hat{L}_i}{\vec{e}_j \vdot \hat{\vec{e}}'_l} = i \hbar \epsilon _{ijk} \vec{e}_k \vdot \hat{\vec{e}}'_l .
\end{equation}
More importantly, with the definition 
\begin{equation}
  \hat{L}'_i \equiv \hat{\vec{e}}'_i \vdot \hat{\vb*{L}} ,
\end{equation}
we have the following relations, 
\begin{align}
  \label{eq:commLiLm}\comm{\hat{L}_i}{\hat{L}'_j} &= 0 , \\
  \comm{\hat{L}'_i}{\hat{L}'_j} &= -i \hbar \epsilon_{ijk} \hat{L}'_k .
\end{align}
Given the eigenstates \(\ket{LMK}\) of \(\{\hat{\vb*{L}}^2, \hat{L}_3, \hat{L}'_3\}\), the eigenfunctions are
\begin{align}
  \label{eq:Dfunc-1} &\bra{\{\vec{e}_i\}} \ket{LMK} = \bra{\{\hat{R} \vec{e}'_i\}} \ket{LMK} \\
  \label{eq:Dfunc-2}&= \bra{\{\vec{e}'_i\}} \hat{D} \ket{LMK} \\
  \label{eq:Dfunc-3}&= \sum _{M' K'} \bra{\{\vec{e}'_i\}} \ket{L M' K'} \bra{L M' K'} \hat{D} \ket{LMK} \\
  \label{eq:Dfunc-4}&= \sum _{M'} \bra{\{\vec{e}'_i\}} \ket{L M' K} \bra{L M' K} \hat{D} \ket{LMK} \\
  \label{eq:Dfunc-5}&= \sum _{M'} \bra{\{\vec{e}'_i\}} \ket{L M' K} \bra{L M' K} \hat{D} \ket{LMK} \delta_{M' K} \\
  \label{eq:Dfunc-6}&= \bra{\{\vec{e}'_i\}} \ket{L K K} \bra{L K K} \hat{D} \ket{LMK} \\
  \label{eq:Dfunc-7}&= \bra{\{\vec{e}'_i\}} \ket{L K K} \bra{L K} \hat{D} \ket{LM} . 
\end{align}
From \cref{eq:Dfunc-3} to \cref{eq:Dfunc-4}, we have used the communication relations \cref{eq:commLiLm}. The \cref{eq:Dfunc-5} is valid because 
\begin{align}
  \notag \bra{L M' K} \hat{L}_3 \hat{D} \ket{LMK} &= M' \bra{L M' K} \hat{D} \ket{LMK} \\
  \notag &= \bra{L M' K} \hat{D} \hat{D}^{\dagger} \hat{L}_3 \hat{D} \ket{LMK} \\
  \notag &= \bra{L M' K} \hat{D} \hat{L}'_3 \ket{LMK} \\
  &= K \bra{L M' K} \hat{D} \ket{LMK}, 
\end{align}
i.e. 
\begin{equation}
  \bra{L M' K} \hat{D} \ket{LMK} = \bra{L M' K} \hat{D} \ket{LMK} \delta _{M' K}.
\end{equation}
When the rotations are described by Euler angles, 
\begin{equation}
  \bra{L K} \hat{D} \ket{LM} = D^{(L)}_{KM} (\alpha,\beta,\gamma),
\end{equation}
and \cref{eq:Dfunc-7} gives us the eigenfunction \cref{eq:15}.

\section{\label{app:b}convergence tests}
\setcounter{equation}{0}
\renewcommand\theequation{B.\arabic{equation}}
\FloatBarrier
\begin{figure}[!h]
  \vspace{-1em}
  \centering
  \subfloat[The ground states.]{
      \centering
      \label{fig:gs_fidelity}
      \includegraphics[width=.46\columnwidth]{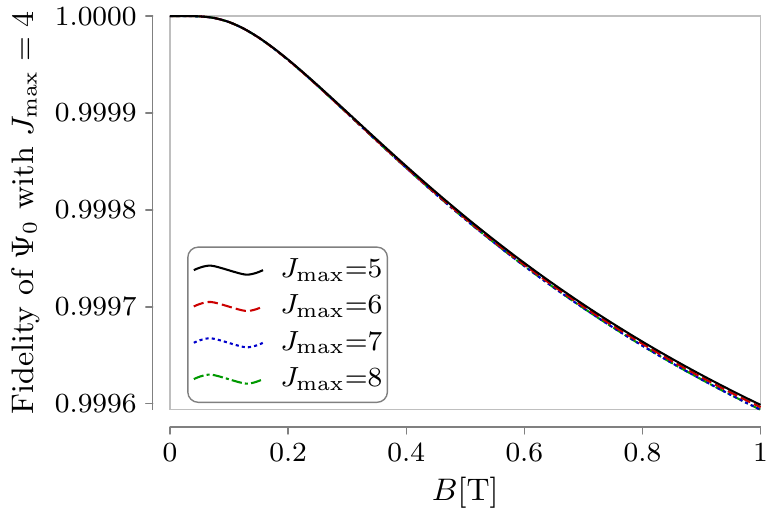}
  }
  \subfloat[The first excited states.]{
      \centering
      \label{fig:1e_fidelity}
      \includegraphics[width=.46\columnwidth]{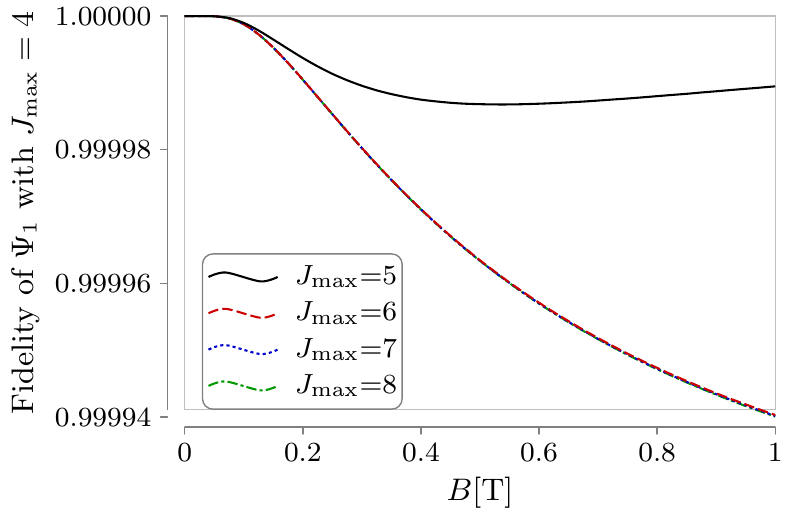}
  }
  \vspace*{-1ex}
      \subfloat[Pure states.]{
          \centering
      \label{fig:entro_L8}
      \includegraphics[width=.46\columnwidth]{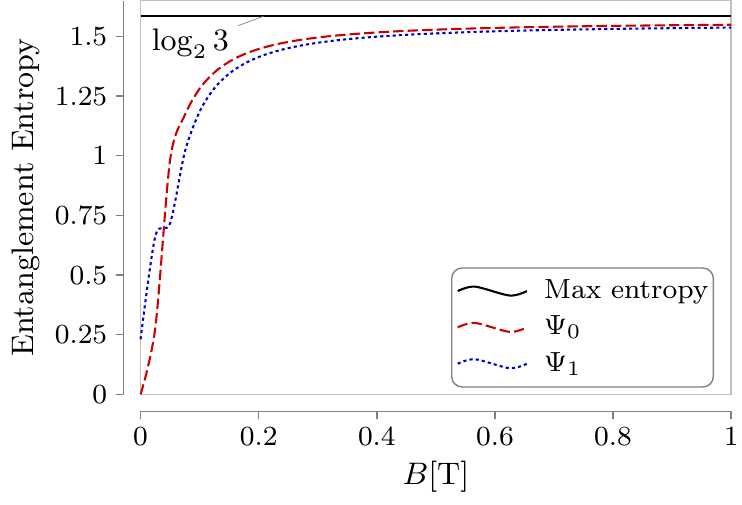}
  }
      \subfloat[Thermal equilibrium states.]{
          \centering
      \label{fig:neg_Lmax2-6}
      \includegraphics[width=.46\columnwidth]{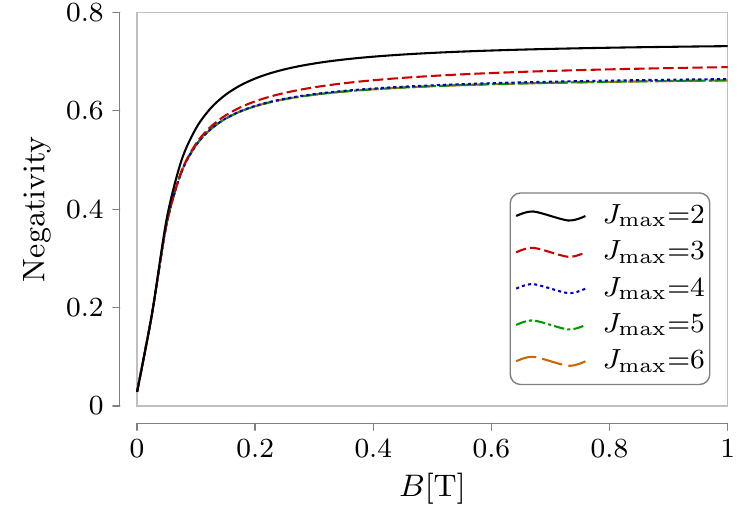}
  }
  \caption[]{\label{fig:fidelity} The fidelity between the \subref{fig:gs_fidelity} ground states (\subref{fig:1e_fidelity} 1-st excited states) of \(J_\mathrm{max} = 4\) and \(J_\mathrm{max} \in \{5,6,7,8\}\). \subref{fig:entro_L8} The entanglement entropy of the ground state and the 1-st excited state with cutoff \(J_{\mathrm{max}} = 8\). \subref{fig:neg_Lmax2-6} The negativity of the thermal equilibrium states with different cutoff \(J_{\mathrm{max}} \in \{2,3,4,5,6\}\) at temperature \(T = \SI[]{10}{mK}\).}
\end{figure}

In this appendix, we display the convergence tests of the cutoff of maximum angular momentum \(J_{\mathrm{max}}\) in our numerical calculation.  

Fidelity is a measure of the "closeness" of two quantum states and is defined as the quantity 
\begin{equation}
  F(\rho , \sigma) = \biggl(\mathrm{tr} \sqrt{\sqrt{\rho} \sigma \sqrt{\rho}} \biggr)^2.
\end{equation}
In the special case where \(\rho\) and \(\sigma\) are pure quantum sates, namely, \(\rho = | \psi_{\rho} \rangle \langle \psi_{\rho} |\) and \(\sigma = | \psi_{\sigma} \rangle \langle \psi_{\sigma} |\), the definition reduces to the squared overlap between the states: 
\begin{equation}
    F(\rho , \sigma) = | \langle \psi_{\rho} | \psi_{\sigma} \rangle |^2.
\end{equation}
We check the fidelity between the ground states of \(J_\mathrm{max} = 4\) and \(J_\mathrm{max} \in \{5,6,7,8\}\) as shown in \cref{fig:gs_fidelity}, and see that the ground state of larger cutoff (\(J_\mathrm{max} \geq 4\)) changes very little. The case of the 1-st excited state shown in \cref{fig:1e_fidelity} is the same.

As a comparison with \cref{fig:4a} which shows the entanglement entropy with a cutoff of angular momentum \(J_\mathrm{max} = 4\), a cutoff of \(J_\mathrm{max} = 8\) is shown in \cref{fig:entro_L8}. For the thermal equilibrium states, we check their negativity with several cutoff \(J_\mathrm{max} \in \{2,3,4,5,6\}\) at temperature \(T = \SI[]{10}{mK}\) which is shown in \cref{fig:neg_Lmax2-6}. It is clear to see that the negativity is convergent for \(J_\mathrm{max} \geq 4\). 

\section{\label{app:c}Hamiltonian in space-fixed frame}
\setcounter{equation}{0}
\renewcommand\theequation{C.\arabic{equation}}

In this appendix, we give a calculation of the model solved in space-fixed frame $\{\vec{e}_1, \vec{e}_2, \vec{e}_3\}$. A direct calculation shows that the complete set of commuting observables should be \(\{\hat{\vb*{L}}^2, \hat{L}_3, \hat{L}'_3,  \hat{\vb*{S}}^2, \hat{S}_3\}\), and the Hamiltonian is written as 
\begin{align}
  H &= \frac{D}{\hbar} (\sum_{i=1}^{3}\hat{S}_i \vec{e}_i \vdot \vec{e}'_3)^2  + \frac{1}{2} \frac{\hat{\vb*{L}}^2}{I_1} + \frac{1}{2}\biggl( \frac{1}{I_3} - \frac{1}{I_1} \biggr) {\hat{L}_3^{\prime 2}} \notag \\
  &\quad + \frac{g \mu_B}{\hbar} B \hat{S}_3. 
\end{align}
One can solve the eigen problem of this Hamiltonian following the same procedure in the main text. The entanglement is between the NV\(^-\) spin and the mechanical rotation of the nanodiamond which is shown in \cref{fig:5}. 
\FloatBarrier

\bibliography{paper.bib}
\end{document}